\documentclass[prl,twocolumn,superscriptaddress,floatfix,10pt]{revtex4-1}
\usepackage{graphicx,bm}
\usepackage{amsmath}
\usepackage{amsfonts}
\usepackage{amssymb}
\usepackage{braket}
\usepackage{bm}
\usepackage{color}

\usepackage{xcolor}
\usepackage[%
colorlinks=true,
urlcolor=blue,
linkcolor=blue,
citecolor=blue
]{hyperref}

\begin{document}

	\title{Twisted bilayer graphene in a parallel magnetic field}
	\author{Yves H. Kwan}
	\affiliation{Rudolf Peierls Centre for Theoretical Physics,  Clarendon Laboratory, Oxford OX1 3PU, UK}
	\author{S. A. Parameswaran}
	\affiliation{Rudolf Peierls Centre for Theoretical Physics,  Clarendon Laboratory, Oxford OX1 3PU, UK}
	\author{S. L. Sondhi}
	\affiliation{Department of Physics, Princeton University, Princeton, New Jersey 08540, USA}

	\begin{abstract}
		We study the effect of an in-plane magnetic field on the non-interacting dispersion of twisted bilayer graphene. Our analysis is rooted in the chirally symmetric continuum model, whose zero-field band structure hosts exactly flat bands and large energy gaps at the magic angles. At the first magic angle, the central bands respond to a parallel field by forming a quadratic band crossing point (QBCP) at the Moir\'e Brillouin zone center. Over a large range of fields, the dispersion is invariant with an overall scale set by the magnetic field strength. For deviations from the magic angle and for realistic interlayer couplings, the motion and merging of the Dirac points lying near charge neutrality are discussed in the context of the symmetries, and we show that small magnetic fields are able to induce a qualitative change in the energy spectrum.
		We conclude with a discussion on the possible ramifications of our study to the interacting ground states of twisted bilayer graphene systems.
	\end{abstract}
	\date{\today}
	
	\maketitle

	\textit{Introduction.---} 
	Flat band systems allow electron interactions to dominate kinetic energy, providing a favorable setting for exotic correlated and topological phases of matter to emerge~\cite{parameswaran2013,bergholtz2013}. A recent entrant to their roster is twisted bilayer graphene (TBG), a van der Waals heterostructure~\cite{geim2013} that, absent interactions, exhibits Dirac points at charge neutrality with a suppressed Fermi velocity compared to monolayer graphene~\cite{lopes2007,trambly2010,bistritzer2011,lopes2012,sanjose2012,li2010,lucian2011}. The suppression is especially pronounced near small `magic' twist angles at which  the two central Moir\'e bands (per valley and spin) become nearly flat in the Moir\'e Brillouin zone (mBZ)~\cite{bistritzer2011}. 
	The rich possibilities this affords to  correlation effects are strikingly illustrated by the experimental discovery of proximate Mott insulating and superconducting phases~\cite{cao2018a,cao2018b}, and more recent observations of the anomalous Hall effect and orbital ferromagnetism~\cite{sharpe2019,lu2019,serlin2019} in graphene Moir\'e heterostructures. Concomitantly, a large body of theoretical work has emerged to explain various aspects of TBG and related Moir\'e heterostructures~\cite{po2018,zou2018,po2019,kang2018,koshino2018,xu2018,yuan2018,xie2018,tarnopolsky2019,bultinck2019,song2019,lian2019,thomson2018,wu2018,zhang2019a,zhang2019b,liu2019}.
	
	Part of the allure of TBG arises from its remarkable tunability. Due to the small size of the mBZ, gate-induced doping of the flat bands  is relatively easy to achieve. Hydrostatic pressure increases the interlayer coupling strength and the magic angles~\cite{carr2018,yankowitz2019}. The purely two-dimensional setting allows scanning probes to directly interrogate real-space electronic correlations.
	Here, we consider the effects of another tuning parameter --- an in-plane magnetic field. Although considered in Ref.~\cite{roy2013}, this has been little explored in the flat band setting where, as we show, it plays an enhanced role. Starting from the limit of exactly flat and degenerate bands~\cite{tarnopolsky2019}, we find that even an infinitesimal in-plane field drives the formation of a quadratic band crossing point (QBCP) at the mBZ center. Strikingly, the shape of the resulting dispersion is invariant for experimentally accessible field strengths, which only serve to set the overall bandwidth. For deviations from this flat band limit, we study the motion and merging of Dirac points, and extend our analysis to realistic interlayer couplings, showing that qualitative features survive away from the chiral limit. We also explore the interplay between orbital and spin effects of in-plane fields, and contextualize our findings within the interacting states of TBG. {\it In toto}, our results suggest that parallel fields offer an intriguing new `knob' to explore this intriguing class of Moir\'e materials.

	\textit{Twisted bilayer geometry \& continuum model.---} 
	\begin{figure}[]
		\includegraphics[trim={0.3cm 15.9cm 0.8cm 0cm}, width=1\linewidth,clip=true]{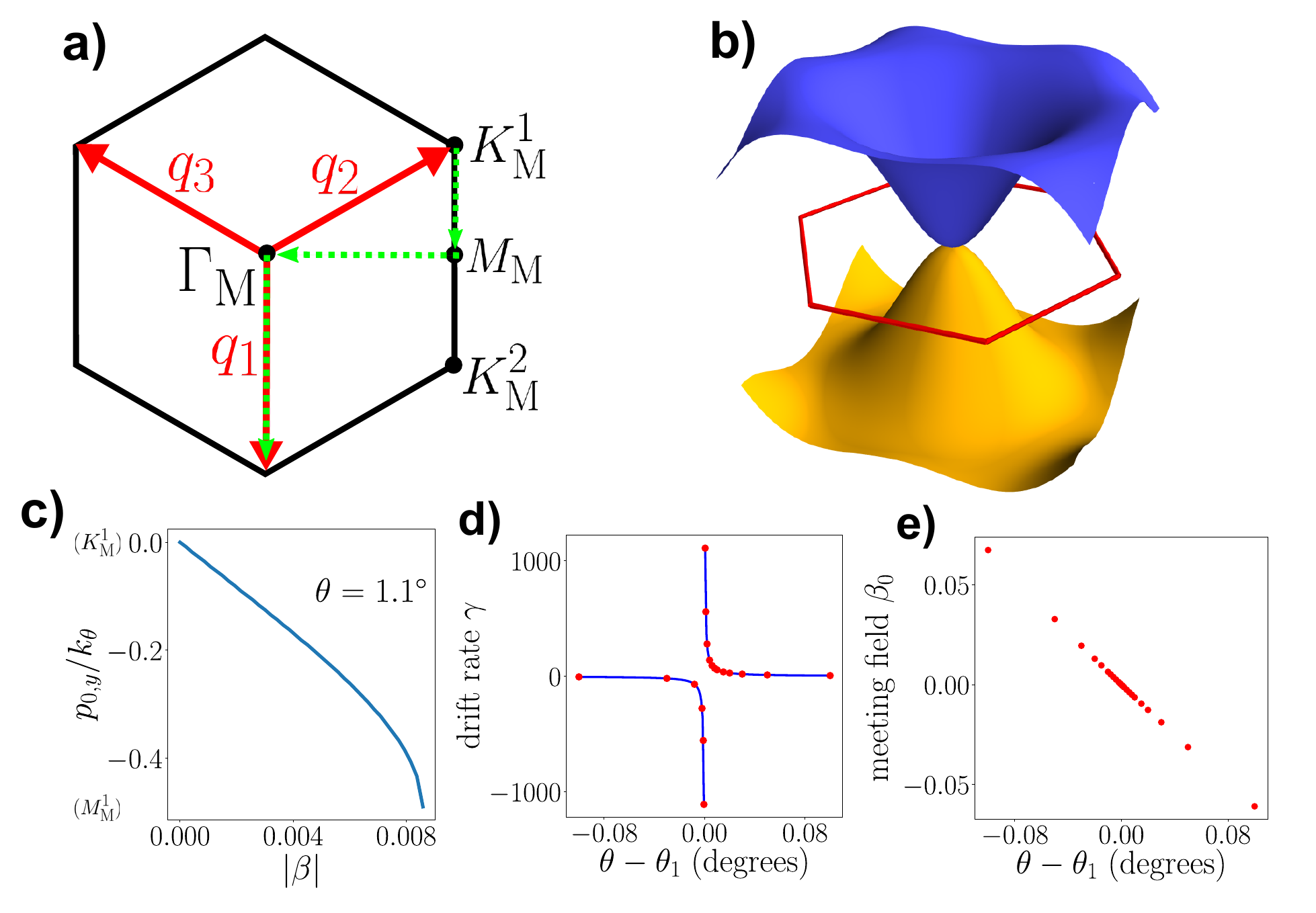}
		\caption{(a) Moir\'e Brillouin zone.
		The Moir\'e  and Dirac wavevectors are $k_\theta=2k_D\sin(\theta/2)$ and
		$k_D=\frac{4\pi}{3\sqrt{3}a}$ with $a=1.42$\AA~ the C--C bond length. Green lines show the mBZ path used for Fig.~\ref{Figsnakepath}. (b) Central bands at $\theta_1$ in a parallel magnetic field, showing quadratic band crossing point (QBCP) at $\Gamma_\text{M}$. (c) The position of the Dirac point originally at $K^1_\text{M}$ as the dimensionless magnetic field is increased along the easy direction. (d) The initial dimensionless drift rate of the Dirac point as a function of angle is computed (red dots) and compared to perturbation theory (blue line).
		(e) Field along $\hat{x}$ required to induce Dirac point merging as a function of angle.}
		\label{Figdiracpointpositions}
	\end{figure}
	Consider AA-stacked bilayer graphene with interlayer spacing $c_0$~=~3.35~\AA. Layers 1 and 2 are placed at $z=\pm c_0/2$ and rotated by $\pm\theta/2$ respectively about a hexagon center; this choice preserves the maximal subset of monolayer symmetries including $C_{\text {6v}}$ rotation about the hexagon center~\cite{zou2018}.
	Generic twist angles lead to an incommensurate structure. 
	However, the ``continuum approximation''~\cite{bistritzer2011} yields a Hamiltonian that is periodic with Moir\'e wavector $k_\theta=2k_D\sin\theta/2$~\footnote{The emergent periodicity arises because the intralayer physics is modelled by Dirac terms, which is a good approximation when $|\bm{k}-\bm{K}|a\ll1$ (hence the ``continuum'' designation). If we include the full monolayer dispersion, this permits purely in-plane Bragg scattering by a graphene reciprocal lattice vector $\bm{b}_j$. In general, $\bm{b}_j$ may not
	 be expressible as an integer sum of Moir\'e reciprocal lattice vectors $\bm{b}_{\text{M},j}$, which is equivalent to incommensurability.}, and furthermore promotes the approximate valley $U(1)$ to an exact symmetry. 
	Restricting to states near monolayer valley $K$, the intralayer physics is captured by Dirac dispersions centered at $\bm{K}^{1,2}\equiv R_{\pm\theta/2}\bm{K}$.
	These momenta fold onto mBZ corners (Fig.~\ref{Figdiracpointpositions}) upon including interlayer coupling
 in the dominant harmonic approximation~\cite{bistritzer2011}. In the Bloch spinor basis \mbox{$\psi(\bm{p})=[\psi_{1A}(\bm{p}),\psi_{1B}(\bm{p}),\psi_{2A}(\bm{p}),\psi_{2B}(\bm{p})]^T$} with $\bm{p}$ measured relative to the Dirac point the Hamiltonian is
	\begin{equation}\label{EqnContinuumModel}
	\begin{gathered}
	H(\bm{p},\bm{p}')=
	\begin{pmatrix}
	\hbar v_0 \bm{\sigma}^*_{\theta/2}\cdot\bm{p}\,\delta_{\bm{p},\bm{p}'} &
	\sum_{i=j}^3T_j\delta_{\bm{p}-\bm{p}',\bm{q}_j}\\
	\sum_{i=j}^3T_j\delta_{\bm{p}-\bm{p}',-\bm{q}_j} &
	\hbar v_0 \bm{\sigma}^*_{-\theta/2}\cdot\bm{p}\,\delta_{\bm{p},\bm{p}'}
	\end{pmatrix}\\
	\end{gathered}
	\end{equation}
	where  $v_0$~=~8.8~$\times$~10$^{\text 5}$~ms$^{-\text 1}$ is the bare Fermi velocity, the Pauli matrices act in sublattice space,
	$\bm{\sigma}_{\theta/2}=e^{i\theta\sigma_z/4}\bm{\sigma}e^{-i\theta\sigma_z/4}$,   
	$T_j = w_{AA} \sigma^0 + w_{AB}(\cos\frac{2\pi (j-1)}3 \sigma^x + \sin\frac{2\pi (j-1)}3 \sigma^y)$,  and the  $\bm{q}_j$ are given in Fig.~\ref{Figdiracpointpositions}(a). 
	$w_{AA}$ ($w_{AB}$) parameterizes the AA/BB (AB/BA) interlayer hopping strengths, and $w_{AA}< w_{AB}$
	due to lattice relaxation effects~\cite{nam2017,carr2019}. The interlayer coupling leads to downward renormalization of the Fermi velocity, especially near the magic angles $\{\theta_m\}$.
	
	For $w_{AA}=0$~\cite{tarnopolsky2019}, the model acquires a chiral/particle-hole symmetry because \mbox{$\{H,\sigma_z\}=0$}. Its spectrum depends only on the dimensionless angle parameter $\alpha\equiv w_{AB}/v_0\hbar k_\theta$, where $w_{AB}=0.11$~eV. 
	Remarkably, the two central bands become degenerate, exactly flat, and energetically isolated across the mBZ at the magic angles. 
	We will initially focus on this idealized limit, and consider twists near the first magic angle $\theta_1\simeq1.086^\circ$.

	\begin{figure*}[t!]
		\includegraphics[trim={-2cm 22.2cm -2.2cm 0cm}, width=1\linewidth,clip=true]{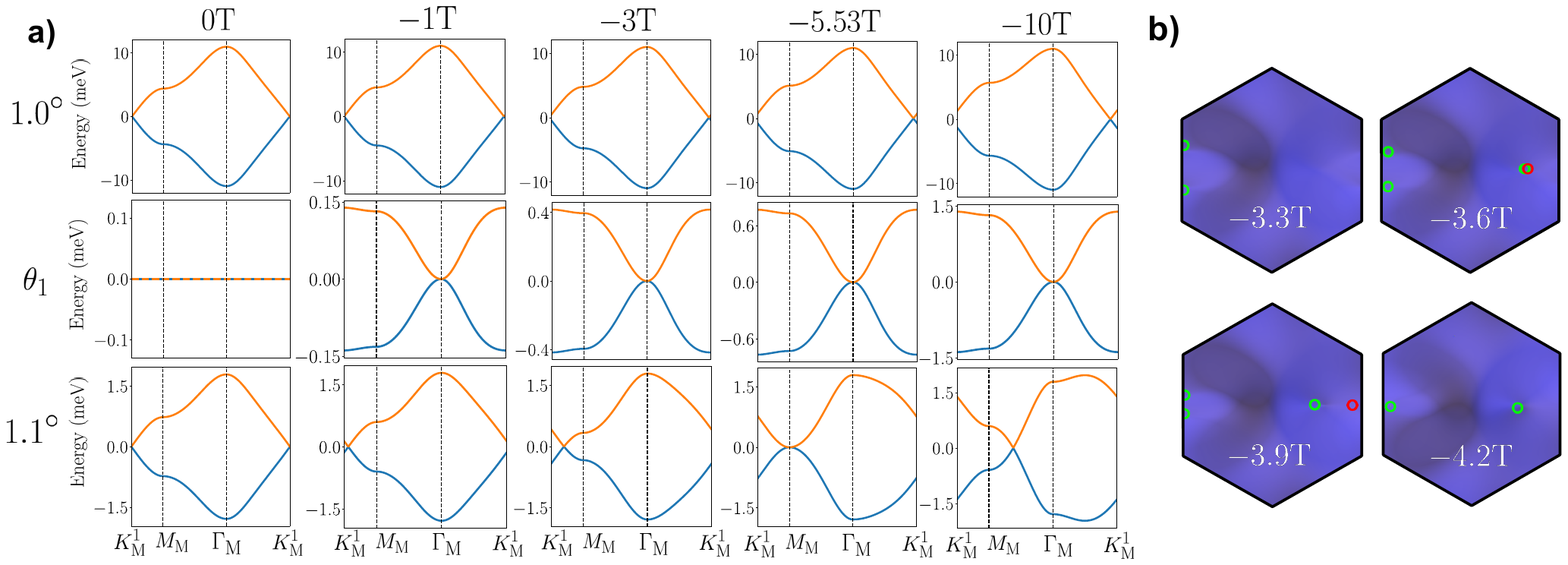}
		\caption{(a) The central bands of the chirally symmetric model ($w_{AA}=0$) along $K^1_\text{M}\rightarrow M_\text{M}\rightarrow\Gamma_\text{M}\rightarrow K^1_\text{M}$ [see Fig.~\ref{Figdiracpointpositions}(a)].
			The nearest bands are separated by $\sim100$~meV. 
			(b)~Evolution of  Dirac points in the mBZ (black hexagon), for $w_{AA}=0.06$~eV and $\theta=1.1^\circ$. Green/red circles correspond to $\pm1$ winding number. 
		}
		\label{Figsnakepath}
	\end{figure*}
	
	\textit{In-plane  field.---}
	Orbital effects of an in-plane magnetic field $\bm{B}=B(\cos\chi,\sin\chi,0)$ can be incorporated by minimally coupling kinetic terms to a vector potential $\bm{A}=z\bm{B}\times\bm{\hat{z}}$ chosen to
 preserve the Moir\'e periodicity: 
	\begin{equation}\label{EqnKineticMinimalCoupling}
	\begin{aligned}
	& \bm{\sigma}^*_{\pm\theta/2}\cdot{\bm{p}} \mapsto \bm{\sigma}^*_{\pm\theta/2}\cdot\left[\bm{k}-\bm{K}^{l}\pm\frac{eBc_0}{2\hbar}\begin{pmatrix}\sin\chi\\-\cos\chi\end{pmatrix}\right]\\
	\end{aligned}
	\end{equation}
where we take upper (lower) signs for layer $l=1$ ($l=2$).
	
	Symmetries mandate that the system remains semimetallic for nonzero  magnetic field. 
	For $\bm{B} =0$ TBG hosts two Dirac points which are pinned at the mBZ corners $K^{1,2}_\text{M}$ due to $\hat{C}_{\text{3v}}$ symmetry. Combined time-reversal (TRS) and two-fold rotation $\hat{C}_{\text{2v}}\mathcal{\hat{T}}$ (realized here as  $\sigma_x\hat{\mathcal{K}}\{\bm{r}\rightarrow-\bm{r}\}$, where $\hat{\mathcal{K}}$ is complex conjugation) quantizes the Berry phase 
	to 0 or $\pi$, endowing the Dirac points with a $\mathbb{Z}_2$ charge and protecting them from gapping out individually~\cite{goerbig2017}. A stronger condition holds as long as the central pair of bands are energetically isolated: because the Dirac points arise from the same valley, they can be assigned identical $\mathbb{Z}$ winding numbers which are protected in a two-band model;  hence they cannot annihilate unless they touch other bands~\cite{po2018,gail2011,goerbig2017,ahn2019}.
	
	An in-plane magnetic field preserves $\hat{C}_{\text{2v}}\mathcal{\hat{T}}$, since $\hat{C}_{\text{2v}}$ and $\mathcal{\hat{T}}$ each flip the field direction. 
	Furthermore, chiral symmetry is preserved so the Dirac points remain at charge neutrality. However, any field orientation breaks $\hat{C}_{\text{3v}}$, meaning the Dirac points are no longer pinned to the mBZ corners. Therefore we expect them to drift around the mBZ as the magnetic field is ramped up. Since TRS is explicitly broken, the band structure for the $K$ and $K'$ valleys are no longer time-reversed versions of each other.
	
	Intuitively, we can understand this drift in the limit of uncoupled layers~\footnote{More precisely we consider the limit of infinitesimal interlayer coupling, so that the gauges in each layer cannot be separately transformed.}: from Eq.~\eqref{EqnKineticMinimalCoupling} we read off that the Dirac points of two layers are shifted in opposite directions by $\delta\bm{k} \propto B$. 
	Since they cannot annihilate, it is natural to expect the most interesting effects when they meet. This is easiest to arrange for the field orientation $\bm{B}=B\hat{x}$, where for $B<0$ the Dirac points move towards  (away) from each other along the $\hat{k}_y$ axis in valley $K$ ($K'$). For decoupled layers, the fields required are prohibitively large. However, a simple estimate yields a characteristic field energy scale of $ec_0 v_0\sim 0.3~\rm{meV}\cdot\rm{T}^{-1}$, indicating that in the flat band limit where the competing energy scale (i.e., the bandwidth) is small, the relative impact of a parallel field can be significant.
	
	\textit{Behavior at the first magic angle.---}
	At the first magic angle $\theta_1$, the central bands of chirally-symmetric TBG are exactly flat and degenerate across the mBZ. As soon as a finite field along any in-plane direction is applied, the bands gap and disperse across the entire mBZ except for a QBCP at $\Gamma_\text{M}$ which lies at charge neutrality [Fig.~\ref{Figdiracpointpositions}(b) --- i.e., the two incipient Dirac points immediately migrate to the center of the mBZ. Strikingly, the shape of the spectrum is invariant over a remarkably large range of $B$, which simply sets the overall energy scale [Fig.~\ref{Figsnakepath}(a)].

	We can understand this behavior using degenerate perturbation theory, which gives the energy shift to linear order in $B$. The lack of a competing kinetic energy explains the persistence of the shape of the spectrum. That the bands touch at $\Gamma_\text{M}$ can be understood using point-group symmetry, as follows.
	The central bands at $\Gamma_\text{M}$ (which enjoys all single-valley symmetries) transform as singlets. However, the perturbation in (\ref{EqnKineticMinimalCoupling}) transforms as the 2D irreducible representation. Therefore the energies at $\Gamma_\text{M}$ are unchanged to $O(B)$. The band touching is quadratic since there needs to be a net $2\pi$ pseudospin winding for it to split into two Dirac points, and the chiral symmetry rules out a trivial linearly dispersing term.
	
	Magnetic field scales $eBc_0v_0$  comparable to the $\sim100$~meV gap to the nearest bands (corresponding to  $B\gtrsim 100$~T) allow significant interband mixing. This violates the  emergent protection of the QBCP, allowing it to split. For finite $w_{AA}$ or $|\theta-\theta_1|$, the low-field band evolution is no longer shape-invariant, since the $B=0$ bandwidth $W$ provides a new scale. However for small deviations from the exactly flat limit, the spectrum recovers an approximate QBCP for $B\gg W/ec_0 v_0\sim 1.5~\rm{T}$ for typical samples at the magic angle where $W\sim 5~\rm{meV}$.

	\textit{Dirac point motion away from $\theta_1$.---}
	Fig.~\ref{Figsnakepath}(a) displays the evolution of the band structure of chiral TBG close to $\theta_1$ with changing $\bm{B}=B\hat{x}$. The path in the mBZ [Fig.~\ref{Figdiracpointpositions}(a)] has been chosen to track the Dirac point originally at $K^1_\text{M}$. The location of the other Dirac point
	is determined since two-fold rotation about the $x$-axis is a symmetry. Actually for vanishing $w_{AA}$ and $\bm{B}\parallel\hat{x}$, we note that there is another symmetry which demands that \mbox{$E(k_x,k_y)=-E(-k_x,k_y)$}, where $\bm{k}$ is measured from $\Gamma_\text{M}$. This is because the spinor part of the kinetic terms can be unrotated for $w_{AA}=0$ by a unitary transform, leading to a particle-hole symmetry operator of the form $\sigma_y\{k_x\rightarrow-k_x\}$. Taken together, these imply that the spectrum is invariant under reflection in $k_x$ and $k_y$, strongly constraining the motion of the Dirac points.
	
	The Dirac points drift along the $k_y$ axis for small magnetic fields. As they approach each other, the associated van Hove singularities appear at lower energy. However there are several key differences to the uncoupled layer case. The direction in which the Dirac points move changes sign across $\theta_1$, and the magnitude of the drift varies non-linearly with $B$. Furthermore, the rate of drift with $B$ is greater for angles close to $\theta_1$, and appears to diverge as the magic angle is approached. 
	
	When the Dirac points meet at $M_\text{M}$, they form a QBCP instead of gapping out.
	When $|B|$ is increased further, the QBCP splits into two Dirac points that move along $\hat{k}_x$ and drift towards $\Gamma_\text{M}$. This merging transition of Dirac points with the same winding has been discussed previously in the context of rotational and stacking defaults in graphene-based systems~\cite{gail2012a,gail2012b,montambaux2018}. 
	The Dirac points do not reach $\Gamma_\text{M}$, even for large $|B|$ and instead eventually reverse direction. Their point of closest approach can be roughly estimated in terms of the energy scale of the $B=0$ gap at $\Gamma_{\rm M}$.
	
	If the magnetic field is opposite to the `easy' direction, for example the case of $\theta<\theta_1$ and $B<0$ as shown in the top row in Fig.~\ref{Figsnakepath}(a), the Dirac points drift vertically towards $\Gamma_\text{M}$ but do not reach it. 
	Note that the situation in valley $K'$ is time-reversed.

	\textit{Perturbation theory about $K^1_\text{M}$.---}
	Analytical predictions for the Dirac point motion can be obtained by considering a truncated version of Eq.~\eqref{EqnContinuumModel}. Within the effective 8-band model of Ref.~\cite{bistritzer2011}, perturbation theory predicts that the Dirac points move a distance $\beta v_0k_\theta/2v^*$ for small dimensionless magnetic fields \mbox{$\beta=eBc_0/\hbar k_\theta$}, 
	where \mbox{$v^*=\frac{1-3\alpha^2}{1+3\alpha^2}v_0$} is the field-independent renormalized Fermi velocity. 
	We find that the features of Fig.~\ref{Figsnakepath}(a), namely the direction and magnitude of Dirac point drift near $\theta_1$ for small $\beta$, are exhibited in this simple calculation, which predicts that the drift rate with $\beta$ diverges as $\sim1/|\theta-\theta_1|$. 
	Fig.~\ref{Figdiracpointpositions}(c) charts the position of the Dirac point as a function of $\beta$ for $\theta=1.1^\circ$, showing that the initial drift is indeed linear;  Fig.~\ref{Figdiracpointpositions}(d) shows the drift rate \mbox{$\gamma(\theta)\equiv\frac{\partial (p_0/k_\theta)}{\partial\beta}\big|_{\beta=0}$} as a function of angle which agrees very well with perturbation theory.

	\textit{Perturbation theory about $M_\text{M}$.---}
	A similar 4-band model can be used to analyze the merging transition at $M_\text{M}$, and predicts that the Dirac points meet at a field $\beta_0\sim|\theta-\theta_1|$ --- the linear scaling agrees well with the numerics in Fig.~\ref{Figdiracpointpositions}(e). 		
	At Dirac point coincidence, the perturbative effect of $\bm{p}$ and $\delta\equiv\beta-\beta_0$ can be captured
	~\cite{mccann2006} via the low-energy effective Hamiltonian 
	\begin{equation}
	H_\text{eff}=\frac{1}{2\alpha}\left[\left(\alpha\delta+\frac{p_x^2-p_y^2}{1+\frac{\delta}{4\alpha}}\right)\sigma_x+\frac{2p_xp_y}{1+\frac{\delta}{4\alpha}}\sigma_y\right]
	\end{equation}
	which matches the form of the ``universal Hamiltonian'' describing merging of two Dirac points with the same winding number~\cite{montambaux2018}. Therefore this 4-band model captures the `right-angle turning' of the Dirac points, and also the square-root scaling with $\delta$ as seen in Fig.~\ref{Figdiracpointpositions}(c).

	\textit{Finite AA coupling.---} 
	The chirally symmetric model is an artificial limit of TBG, and a natural question is whether the phenomenology outlined above persists for more realistic values of $w_{AA}$.
	Fig.~\ref{Figsnakepath}(b) shows the drift of Dirac points at $\theta=1.1^\circ$ with $w_{AA}=0.06$~eV. While the chirally symmetric model has a dispersion invariant under $k_x\rightarrow -k_x$, this no longer holds with finite $w_{AA}$. Another difference is the lack of particle-hole symmetry, which means that the band touchings now lie at different energies. The Dirac points initially drift towards each other roughly along the $k_y$ axis. 
	At larger fields, a series of annihilation/creation events of Dirac points of opposite windings occurs.
	A key distinction here is that in the limit where the Dirac points are close to $\Gamma_\text{M}$, the lack of chiral symmetry permits a term  $\propto \bm{B}\cdot \bm{k}$ to `tilt' the approximate QBCP.   Thus, while the details are  more involved in the chiral limit,  parallel fields can qualitatively modify the the band structure even for realistic $w_{AA}$. 
	
	\textit{Discussion.---} 
	\begin{figure}[t!]
		\includegraphics[trim={0.7cm 23.5cm 3cm 0cm}, width=1\linewidth,clip=true]{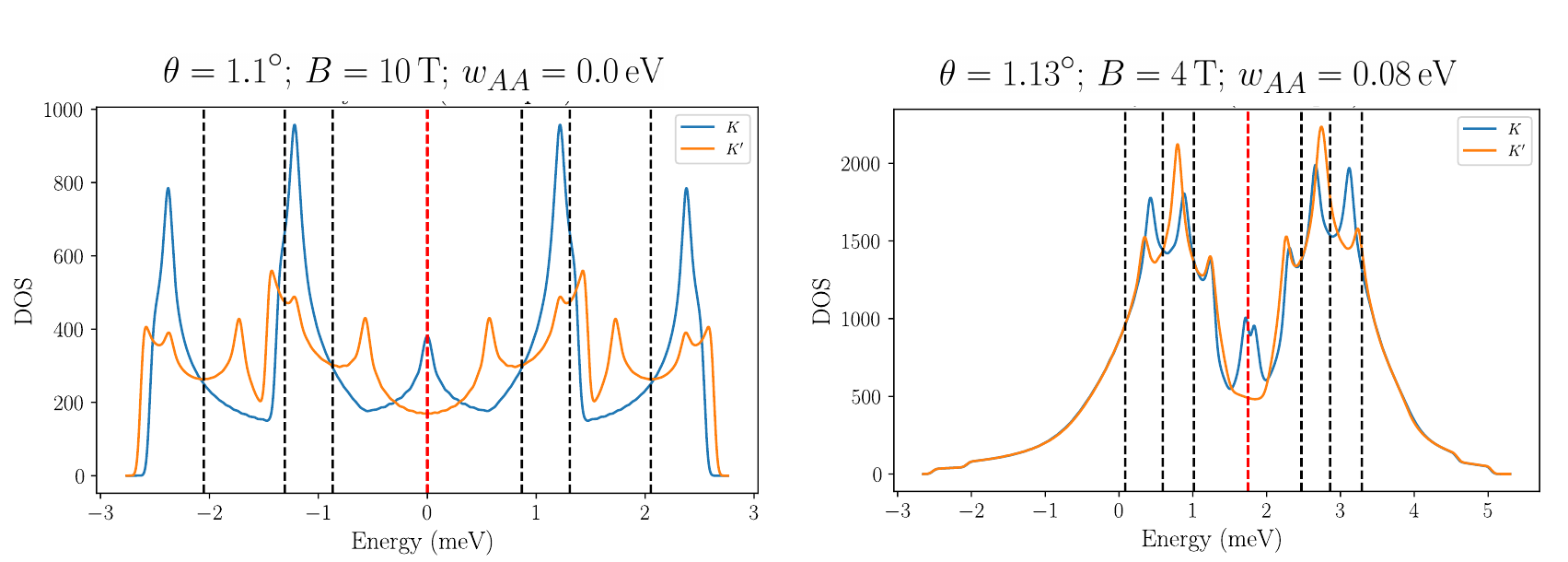}
		\caption{Valley-resolved DOS of the central bands, including both orbital and spin effects. The magnetic field direction corresponds to the `easy' direction for valley $K$. Red dashed line indicates charge neutrality, while black dashed line indicates other integer fillings of the central bands (8 in total).}
		\label{Figdensityofstates}
	\end{figure}
	The in-plane field acts asymmetrically on the two valleys. There will also be further splitting $\Delta E_s = g_s\mu_BB$ due to the spin Zeeman effect, where the spin g-factor $g_s\simeq2$ for graphene systems~\cite{song2010,kurganova2011}. These four flavors should be considered together to get a fuller picture of the effect of $B$. A useful diagnostic here is the total density of states (DOS). 
	The most significant orbital effect is to move Dirac points and tune the energy of their van Hove singularities. If these energies coincide with the Zeeman splitting, then valley-resolved features can appear in the DOS at charge neutrality. This is shown in Fig.~\ref{Figdensityofstates} for both $w_{AA}=0$ and $w_{AA}=0.08~\rm{eV}$, where a peak in the DOS for one of the valleys is apparent for $B\leq10~\rm{T}$. The shuffling of van Hove singularities is expected to have a direct impact on the metallic states and on their correlation-driven instabilities.

\begin{figure}[t!]
		\includegraphics[trim={-0.5cm 21cm 6cm 0cm}, width=1\linewidth,clip=true]{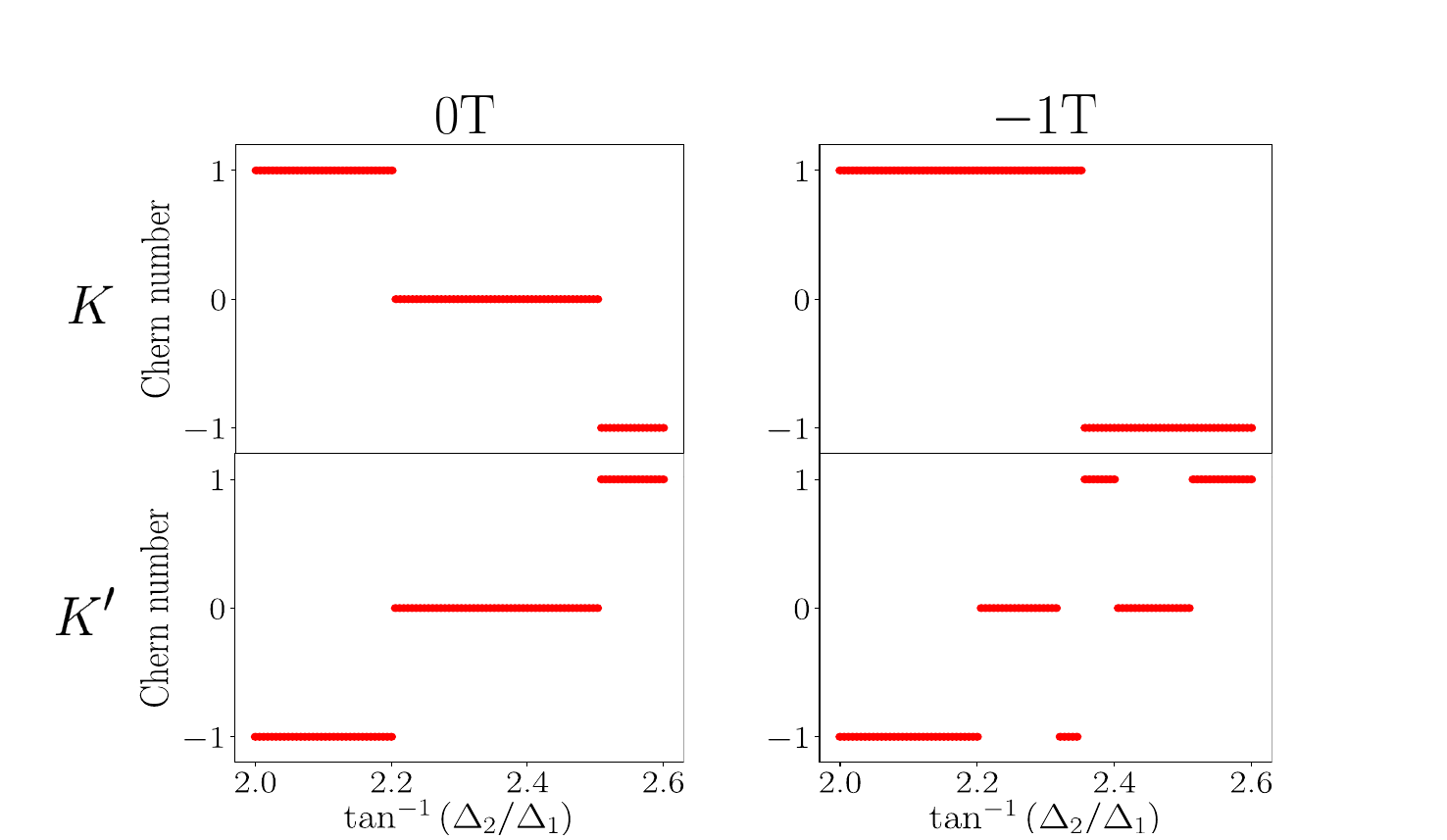}
		\caption{Conduction band Chern number for $w_{AA}=0.08~\rm{meV}$ and $\theta=1.1^\circ$, as a function of substrate configuration --- the quantity \mbox{$\sqrt{\Delta_1^2+\Delta_2^2}=10$~meV} is held constant.
		}
		\label{Figchernnumber}
	\end{figure}

	Many of the most intriguing aspects of TBG lie in the properties of its various correlated ground states.
	One class of such correlated insulators emerges after the opening of single-particle gaps due to interactions within the resulting Chern bands. For example, a partially aligned hexagonal boron nitride (hBN) substrate~\cite{sharpe2019,serlin2019,jung2015} on the top/bottom layer of TBG can be modeled by a uniform perturbation $\sim\Delta_{1(2)}\sigma_z$. This explicitly breaks $\hat{C}_{\text{2v}}$ and therefore gaps out the Dirac points --- the resulting bands have Chern numbers $C=0,\pm1$ depending on the sublattice-splitting strengths $\Delta_{1(2)}$~\cite{bultinck2019,zhang2019b}. Because of TRS, the Chern number configurations for the two valleys are opposite, which factors into the energetic balance between valley-polarized and intervalley coherent ground states at integer filling~\cite{bultinck2019,zhang2019a,zhang2019b}. However, the magnetic field violates TRS, removing the constraints on $C$ for the different valleys. 
	This is demonstrated in Fig.~\ref{Figchernnumber}, which shows that it is possible in principle to engineer a situation where one valley has $C=0$ while the other has $|C|=1$. Given the recent observation of orbital ferromagnetism and the resulting anomalous Hall effect in hBN-TBG, our work suggests that parallel fields can provide a new route to modifying the band topology that underlies these correlated phases.

	Even absent such substrate-induced effects, TBG still exhibits many correlation-driven phenomena. Here we note that {\it prima facie} the energy scales associated with a parallel field ($0.3~\rm{meV}\cdot\rm{T}^{-1}$) compare favorably with the gaps ($0.1-1~\rm{meV}$) reported for various experimentally realized insulating and superconducting states. It is therefore reasonable to imagine that it introduces a new energy scale that can tune the phase boundaries between different correlated phases. A potentially useful feature is that the effect of the parallel field on the DOS is valley- and spin- resolved. We also note that semimetallic behavior and novel correlated phases  have been reported at charge neutrality, which is intriguing in light of the fact that this is where we anticipate the effect of a parallel field to be most dramatic. However, we caution that previous work~\cite{liu2019} has considered the orbital effects of a parallel field on correlated insulating states at charge neutrality within a Hartree-Fock (HF) picture, and argued that these will be small since the HF bandstructure has its smallest gap at the $\Gamma_\text{M}$-point, where the electronic states do not directly couple to the parallel field. We suggest an alternative perspective, where the orbital shuffling of the single-particle states is taken into account {\it before} considering interactions. We defer a detailed exploration of these questions to the future, and here simply note that the delicate interplay of correlations and band structure suggests that the answers are likely subtle and worthy of further  investigation. 
	\begin{acknowledgments}
	\textit{Acknowledgements.---} We thank B.A. Bernevig  for useful discussions. We acknowledge support from the European Research Council (ERC) under the European Union Horizon 2020 Research and Innovation Programme [Grant Agreement No.~804213-TMCS] (SAP, YHK).
	\end{acknowledgments}

	\appendix
	\section{Appendix: Truncated 8-Band and 4-Band Hamiltonians}
	In this appendix, we elaborate on the truncated Hamiltonians that were used to derive analytical predictions of the Dirac point motion. 
	
	\textit{8-band model about $K^1_\text{M}$.---}
	Define the dimensionless magnetic field \mbox{$\beta=eBc_0/\hbar k_\theta$}. For small fields, there is a Dirac point located at $\bm{K}^1+\bm{p}_0(\beta)$ with $\bm{p}_0(\beta)$ parametrically small in $\beta$. Therefore we focus on the plane wave state $\ket{\bm{K}^1+\bm{p},1}$ with $\bm{p}\simeq0$, which directly couples to $\ket{\bm{K}^2+\bm{p}-\bm{q}_1,2},\ket{\bm{K}^2+\bm{p}-\bm{q}_2,2},$ and $\ket{\bm{K}^2+\bm{p}-\bm{q}_3,2}$. The simplest non-trivial approximation includes only these states, thereby defining an effective 8-band model~\cite{bistritzer2011}. For $\bm{p}=0$ and $\beta=0$, the Hamiltonian has two zero energy modes, which are the two degenerate states at the Dirac point. By considering the $\bm{p}$ and $\beta$-dependent terms as perturbations, first-order degenerate perturbation theory predicts that the Dirac point moves to a new position \mbox{$\bm{p}_0(\beta)=(0,\beta v_0k_\theta/2v^*)$}, with a field-independent renormalized Fermi velocity \mbox{$v^*=\frac{1-3\alpha^2}{1+3\alpha^2}v_0$}~\cite{tarnopolsky2019}. The Dirac slope vanishes at $\alpha^*=1/\sqrt{3}$ or an angle of $\theta^*\simeq1.1^\circ$, which is close to $\theta_1\simeq1.086^\circ$. This is consistent with the general observation that perturbation theory accesses the first magic angle relatively well. For relatively flat linear bands, a small perturbation in the dispersion can shift the band crossing by a large distance, which explains why $p_0\sim1/v^*$. In this framework it is natural to interpret the $\theta_1$ case as Dirac points drifting infinitely quickly away from the mBZ corners. The dimensionless drift rate is defined as \mbox{$\gamma(\theta)\equiv\frac{\partial (p_0/k_\theta)}{\partial\beta}\big|_{\beta=0}$}.
	To quantitatively compare this with numerics, we shift the predicted drift rate in $\theta$ to adjust for the slightly different magic angle estimate.
	
	\textit{4-band model about $M_\text{M}$.---}
	A similar approach can be used to analyze the merging transition of Dirac points at $M_\text{M}$, this time restricting attention to  the two momentum states near $M_\text{M}$ with the smallest intralayer kinetic energies, namely $\ket{\bm{K}^1+\bm{m}+\bm{p},1}$ and $\ket{\bm{K}^1+\bm{m}+\bm{p},2}$, where $\bm{m}\equiv k_\theta(0,-1/2)^T$ and $\bm{p}\simeq0$.
	The Dirac points coincide at $M_\text{M}$ for $\beta_0=2\alpha-1$, implying the scaling $\beta_0\sim|\theta-\theta^*|$ where  $\theta^*\simeq
	1.27^\circ$ deviates slightly from the magic angle $\theta_1 = 1.086^\circ$. The latter is unsurprising given the crudeness of the 4-band approximation. 
	
	At Dirac point coincidence, the eigenstates at $M_\text{M}$ have energies $0,0,\pm2\alpha$. We can capture the perturbative effect of $\bm{p}$ and $\delta\equiv\beta-\beta_0$ 
	~\cite{mccann2006} via the low-energy effective Hamiltonian 
	\begin{equation}
	H_\text{eff}=\frac{1}{2\alpha}\left[\left(\alpha\delta+\frac{p_x^2-p_y^2}{1+\frac{\delta}{4\alpha}}\right)\sigma_x+\frac{2p_xp_y}{1+\frac{\delta}{4\alpha}}\sigma_y\right].
	\end{equation}
	This matches the form of the ``universal Hamiltonian'' describing merging of two Dirac points with the same winding number~\cite{montambaux2018}. At $\delta=0$, the two central bands touch parabolically at $M_\text{M}$. At finite $\delta$, the Dirac points lie symmetrically along the $k_x$ or $k_y$ axis depending on the sign of $\delta$, and lie at a distance $\sim\sqrt{\delta}$ from $M_\text{M}$. Therefore this 4-band model captures the `right-angle turning' of the Dirac points, and also the square-root scaling with $\delta$.

	\clearpage

\end{document}